\documentstyle[12pt,psfig]{article}

\begin{document}
\pagenumbering{arabic}
\begin{flushright}
\end{flushright}
\begin{center}
{\Large \bf  Neutron star composition in strong magnetic fields}\\
\vspace{0.5cm}
{G.~Mao\footnote{e-mail: maogj@mail.ihep.ac.cn \hspace{1cm} Tel: 88200974},
V.N.~Kondratyev, and A.~Iwamoto}\\
{\it Japan Atomic Energy Research Institute \\
Tokai, Naka, Ibaraki 319-1195, Japan} \\
\vspace{0.2cm}
{Z.~Li and X.~Wu} \\
{\it China Institute of Atomic Energy \\
 P.O. Box 275 (18), Beijing 102413, P.R. China}\\
\vspace{0.2cm}
{W.~Greiner}\\
{\it Institut f\"{u}r Theoretische Physik der
J. W. Goethe-Universit\"{a}t\\
 Postfach 11 19 32, D-60054 Frankfurt am Main, Germany}\\
\vspace{0.2cm}
{I.N.~Mikhailov}  \\
{\it Bogoliubov Laboratory of Theoretical Physics \\
JINR, Dubna, Russia and CSNSM-IN2P3, Orsay, France}
\end{center}
\begin{abstract}
\begin{sloppypar}
We study the neutron star composition in the presence of a strong
magnetic field. The effects of the anomalous magnetic moments of
both nucleons and electrons are investigated in relativistic mean
field calculations for a $\beta$-equilibrium system. Since
neutrons are fully spin polarized in a  large field, generally
speaking, the proton fraction can never exceed the field free
case.  An extremely strong magnetic field may lead to a pure
neutron matter instead of  a proton-rich matter.

\end{sloppypar}
\end{abstract}
\vspace{0.5cm}
{\bf PACS} number(s): 26.60.+c; 97.10.Ld; 97.60.Jd
\newcounter{cms}
\setlength{\unitlength}{1mm}
\newpage
\begin{sloppypar}
The discovery of pulsars \cite{Hew68} substantially stimulated the study of
neutron stars, a class of extreme objects in astrophysics. Indeed, shortly
after their discovery pulsars were identified as rotating magnetized neutron
stars \cite{Gol68} with a surface magnetic field of $10^{12}$ - $10^{14}$ G
\cite{Mic91}. Recent observations of soft gamma repeaters (SGRs) have confirmed
that they are newly born neutron stars with very large surface magnetic fields
(up to $10^{15}$ G) \cite{Maz79}. Such stars are named as magnetars \cite{Dun92}.
 It is expected that appreciably higher fields could exist in the interior of
 neutron stars although the total strength of internal magnetic fields remains
 unknown. The allowed internal field strength of a star is constrained by the
 scalar virial theorem \cite{Sha83,Lai91}. According to it, the maximum
 interior field strength could reach $\sim 10^{18}$ G.
 Naturally, one may expect considerable influences of such high fields
 on the properties of neutron stars.

 Theoretical investigation of a free electron gas in intense fields related
 to the neutron star crusts \cite{Lai91} and white dwarfs \cite{Suh99a}
 as well as ideal noninteracting neutron-proton-electron (n-p-e) gas for the
 internal properties of neutron stars \cite{Suh99b} has been carried out
 by several authors. The equation of state (EOS) of pure neutron matter under
 large magnetic fields was investigated in Ref. \cite{Vsh94} where the
 strong interaction was taken into account. Theoretical calculations of spin
 polarized isospin asymmetric nuclear matter based on the
 Br\"{u}ckner-Hartree-Fock formalism has been performed in Ref. \cite{Vid02}.
 Analogous calculations employing various mean-field models can be
 seen in Ref. \cite{Jia01}.
 The effects of strong magnetic
 fields on a dense n-p-e system under the beta equilibrium and the charge
 neutrality conditions were studied in Ref. \cite{Cha97}. Extremely large fields up to the proton critical field
 (here the critical field is defined by the condition that the
 particle's cyclotron energy is comparable to its rest mass
 \cite{Lan77}) $B_{c}^{p} \sim 10^{20}$ G and beyond were considered in their
 calculations. It was found that the nuclear matter in beta equilibrium
 practically converts into  stable {\em proton-rich} matter. However, in these
 investigations the effects of the anomalous magnetic moments (AMM) were
 neglected, which may play an important role in the presence of such high fields.
 Recent studies demonstrated \cite{Bro00} that the contributions from the AMM
 of nucleons become significant when the magnetic field $B > 10^{5}B_{c}^{e}$.
 Here $B_{c}^{e}=4.414 \times 10^{13}$ G is the electron critical field.
 Consequently, the softening of the EOS caused by Landau quantization
 \cite{Cha97} is overwhelmed by the stiffening due to the incorporation of the
 nucleon anomalous magnetic moments. But a dramatic increase of the proton
 fraction with the increase of the magnetic field remains. In the study of Ref.
 \cite{Bro00} the AMM of electrons was not taken into account.

 The anomalous magnetic moment of an electron \cite{Sch48} may not be negligible
 due to its relatively large Bohr magneton. Because of the charge neutrality of
 the system, a significant effect on
 electrons will in turn influence protons. Considering that the proton fraction
 is crucial in determining the direct URCA process which leads to the cooling
 of neutron stars \cite{Bog81,Lat91}, in this work we examine the problem
 with the inclusion of the AMM not only for nucleons but also for electrons.
 One may argue that the electron self-energy may not change
 substantially in magnetic fields when high-order terms are taken
 into account. However, systematic incorporation of high-order
 contributions beyond the AMM is not yet clear, which will be the
 topic of our forthcoming works. Here the effects of magnetic
 fields on different particles within the same system are taken
 into account at the equal footing.
 We consider the magnetic fields up to $10^{6}B_{c}^{e}$. Though we do not yet
 know any object  generating such fields, some possibilities have been
 suggested \cite{Dun00}.

 For the n-p-e system in a uniform magnetic field $B$ along the $z$ axis, the
 interacting Lagrangian can be written as \cite{Ser86}
 \begin{eqnarray}
{\cal L} &=& \bar{\psi} \lbrack i\gamma_{\mu}\partial^{\mu}
   - e \frac{1+\tau_{0}}{2} \gamma_{\mu}A^{\mu} -\frac{1}{4}\kappa_{b}\mu_{N}
   \sigma_{\mu\nu}F^{\mu\nu}
  - M_{N} + {\rm g}_{\sigma}\sigma
   - {\rm g}_{\omega}\gamma_{\mu}\omega^{\mu}
    -\frac{1}{2}{\rm g}_{\rho}\gamma_{\mu} \mbox{\boldmath $\tau$}\cdot
    {\bf R}^{\mu} \rbrack \psi \nonumber \\
 && + \bar{\psi}_{e} \lbrack i \gamma_{\mu}\partial^{\mu} - e \gamma_{\mu}
    A^{\mu} - \frac{1}{4}\kappa_{e}\mu_{B}\sigma_{\mu\nu}F^{\mu\nu}
    -m_{e} \rbrack \psi_{e}
  + \frac{1}{2}\partial_{\mu}\sigma \partial^{\mu}\sigma - U(\sigma)
  \nonumber \\
 &&   -\frac{1}{4}\omega_{\mu\nu}\omega^{\mu\nu} + \frac{1}{2}m_{\omega}^{2}
    \omega_{\mu}\omega^{\mu}
  - \frac{1}{4}{\bf R}_{\mu\nu}\cdot{\bf R}^{\mu\nu} + \frac{1}{2}
 m_{\rho}^{2}{\bf R}_{\mu}\cdot {\bf R}^{\mu},
 \end{eqnarray}
 where the conventional notation has been adopted. $U(\sigma)$ is
 the self-interaction part of the scalar field \cite{Bog77}
 \begin{equation}
 U(\sigma)=\frac{1}{2}m_{\sigma}^{2}\sigma^{2}+ \frac{1}{3}b({\rm g}_{\sigma}
 \sigma )^{3} + \frac{1}{4}c({\rm g}_{\sigma}\sigma )^{4}.
 \end{equation}
 $A^{\mu} \equiv (0,0,Bx,0)$ refers to a constant external magnetic field and
 $\sigma^{\mu\nu}=\frac{i}{2} \lbrack \gamma^{\mu}, \gamma^{\nu} \rbrack$.
 $\mu_{N}$ and $\mu_{B}$ are the nuclear magneton of nucleons and Bohr
 magneton of electrons; $\kappa_{p}=3.5856$, $\kappa_{n}=-3.8263$ and
 $\kappa_{e}=\alpha / \pi$ are the coefficients of the AMM for protons,
 neutrons and electrons, respectively. The third set of parameters
 in Table II of Ref. \cite{Gle91} is used as the nucleon coupling strengths.
 It turns out ${\rm g}_{\sigma}=8.7818$, ${\rm g}_{\omega}=8.7116$,
 ${\rm g}_{\rho}=8.4635$, $b{\rm g}_{\sigma}^{3}=27.9060$, $c{\rm g}_{\sigma}
 ^{4}=-14.3989$. This yields a binding energy $B/A=-16.3$ MeV, saturation
 density $\rho_{0}=0.153$ fm$^{-3}$ and bulk symmetry energy $a_{sym}=32.5$
 MeV.

 The general solutions for the Dirac equation with the inclusion of the AMM are
  in the form  of
  $\psi(X) \propto e^{-iEt + ip_{y}y + ip_{z}z}\phi_{p}(p_{y},p_{z},x)$
 for protons and electrons and $\psi(X) \propto e^{-iEt + i{\bf p}\cdot {\bf x}}
 \phi_{n}({\bf p})$ for neutrons. We have derived the concrete
 expressions of the Dirac spinors $\phi_{p}(p_{y},p_{z},x)$
   and $\phi_{n}({\bf p})$ in the chiral representation \cite{MaoPre}.
  The chemical potentials of protons and neutrons are
  defined as
  \begin{eqnarray}
&&  \mu_{p} = \epsilon_{f}^{p} + {\rm g}_{\omega}\omega_{0} + \frac{1}{2}
    {\rm g}_{\rho}R_{0,0}, \\
 && \mu_{n} = \epsilon_{f}^{n} + {\rm g}_{\omega}\omega_{0} - \frac{1}{2}
    {\rm g}_{\rho}R_{0,0}.
  \end{eqnarray}
  They are related to the respective Fermi momenta via following equations:
  \begin{eqnarray}
 && \left( k_{f,\nu,S}^{p} \right)^{2} = \left( \epsilon _{f}^{p} \right)^{2}
    - \left( \sqrt{m^{*2} + 2eB\nu } + S \Delta \right)^{2},
    \label{pferm} \\
 && \left( k_{f,S}^{n} \right) ^{2} = \left( \epsilon_{f}^{n} \right) ^{2}
   - (m^{*} + S\Delta )^{2}.
  \end{eqnarray}
  Here $\Delta = -\frac{1}{2}\kappa_{b}\mu_{N}B$; $ S=\pm 1$ for spin-up and
  spin-down particles. $\nu$ is the quantum number of Landau levels for
  charged particles \cite{Lan77}. The effective nucleon mass $m^{*}=M_{N}
  -{\rm g}_{\sigma}\sigma$. Here $\sigma$, $\omega_{0}$ and $R_{0,0}$
   are the mean
  values of the scalar field, the time-like component of the vector field and
  the time-like isospin 3-component of the vector-isovector field in neutron
  star matter, respectively. They are obtained by solving the non-linear
  equations of the meson fields. The total scalar density and
 baryon density are $\rho_{S}=\rho_{S}^{p} + \rho_{S}^{n}$ and
 $\rho = \rho_{0}^{p} + \rho_{0}^{n}$ respectively, where
 \begin{eqnarray}
&&  \rho_{S}^{p}= \frac{eBm^{*}}{2\pi^{2}}\sum_{S}\sum_{\nu}
    \frac{\sqrt{m^{*2}+2eB\nu}+S\Delta}{\sqrt{m^{*2}+2eB\nu}} \ln
    \left| \frac{k_{f,\nu,S}^{p} + \epsilon_{f}^{p}}{\sqrt{m^{*2}+2eB\nu}
    + S\Delta} \right|, \\
&&  \rho_{S}^{n}=\frac{m^{*}}{4\pi^{2}}\sum_{S}\left[ \epsilon_{f}^{n}
    k_{f,S}^{n} - (m^{*}+S\Delta)^{2} \ln \left| \frac{k_{f,S}^{n}+ \epsilon
    _{f}^{n}}{m^{*}+S\Delta} \right| \right] , \\
&&  \rho_{0}^{p}=\frac{eB}{2\pi^{2}} \sum_{S}\sum_{\nu}k_{f,\nu,S}^{p},
    \label{pden} \\
&&  \rho_{0}^{n} = \frac{1}{2\pi^{2}}\sum_{S}\left[ \frac{1}{3} \left(
    k_{f,S}^{n} \right) ^{3} + \frac{S\Delta}{2} \left( (m^{*}+S\Delta )
    k_{f,S}^{n} + \left( \epsilon_{f}^{n} \right) ^{2} (\arcsin \frac{m^{*}
    + S\Delta}{\epsilon_{f}^{n}} - \frac{\pi}{2}) \right) \right].
 \end{eqnarray}
 The summation of $\nu$ runs up to the largest integer for which $\left( k_{f,\nu,S}
 ^{p} \right)^{2}$ is positive. For spin-up protons $\nu$ starts from 1
 while for spin-down protons 0. It should be pointed out that here the so-called
 spin up and spin down are just relative notes since the wave functions are
 no longer eigenfunctions of 3-component
  spin operator \cite{MaoPre}, mainly attributed
 to the coupling of the spin to the magnetic field. The electrons are assumed
 to move freely in the strong magnetic field. The chemical potential of
 electrons $\mu_{e}=\epsilon_{f}^{e}$. Its relation to the electron Fermi momentum
 as well as the definition of the electron density $\rho_{0}^{e}$
 are the same as given in
 Eqs.~(\ref{pferm}) and (\ref{pden}) except that the corresponding quantities
 are replaced by the electron ones. Numerical calculations are performed under
 the constraints of the charge neutrality $\rho_{0}^{p}=\rho_{0}^{e}$ and
 the $\beta$-equilibrium $\mu_{n}=\mu_{p}+\mu_{e}$ \cite{Gle97}.
 These two constraint
 equations together with three meson equations are solved self-consistently
 in an iterative procedure.

\begin{figure}[htbp]
  \vspace{0cm}
 \hskip  0.0cm \psfig{file=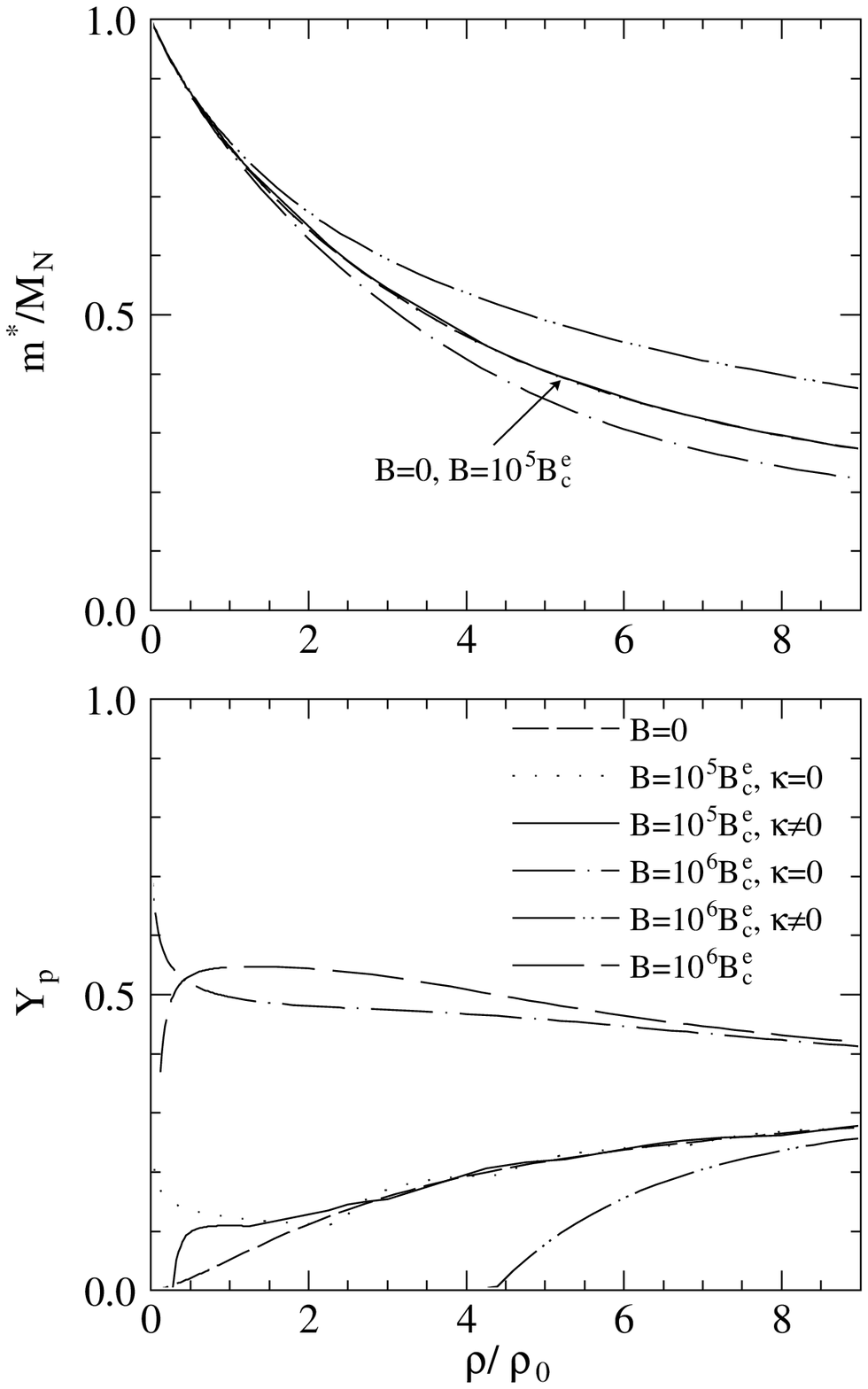,width=14.cm,height=18cm,angle=0}
 \vspace{0.5cm}
 \caption{The effective nucleon mass $m^{*}/M_{N}$ (upper panel) and
 proton fraction Y$_{p}$ (lower panel) as functions of the baryon density.
 Different curves are related to the different cases of magnetic fields
 $B$ and with or without the inclusion of the anomalous magnetic moments
 $\kappa$ as indicated in the figure. In the lower panel, the long-dashed
 line reveals the case considered in Ref. \protect{\cite{Bro00}}. }
 \label{mass}
\end{figure}

 Figure~\ref{mass} displays the effective nucleon mass and proton fraction as
 functions of the baryon density and magnetic field strength. Two situations
 with and without taking into account the anomalous magnetic moments are
 distinguished. In agreement with the finding of Refs. \cite{Cha97,Bro00},
 the effective mass remains unaltered from the field free case when
 $B \leq 10^{5}B_{c}^{e}$. If the field is increased to $10^{6}B_{c}^{e}$,
 the effects of magnetic fields cause the $m^{*}/M_{N}$ to increase when the
 AMM is included ($\kappa \neq 0$) and decrease when the AMM is neglected
 ($\kappa = 0$). The proton fraction is shown to increase with the increasing
 of the magnetic field strength in the case of $\kappa =0$. A considerable
 enhancement of Y$_{p}$ is observed at $B=10^{6}B_{c}^{e}$. For even stronger
 magnetic fields, one may speak about proton-rich matter \cite{Cha97}.
 The situation, however, becomes completely different if the anomalous
 magnetic moments are taken into account.
 At $B=10^{5}B_{c}^{e}$, Y$_{p}$ is a little enhanced at lower density and then
 quickly approaches the field free case as the density increases. In the case
 of $B=10^{6}B_{c}^{e}$, no protons survive at
 $\rho < 4\rho_{0}$. Evident suppression of the proton fraction is
 exhibited for the typical density range of neutron stars.

\begin{figure}[htbp]
  \vspace{0cm}
 \hskip  0.0cm \psfig{file=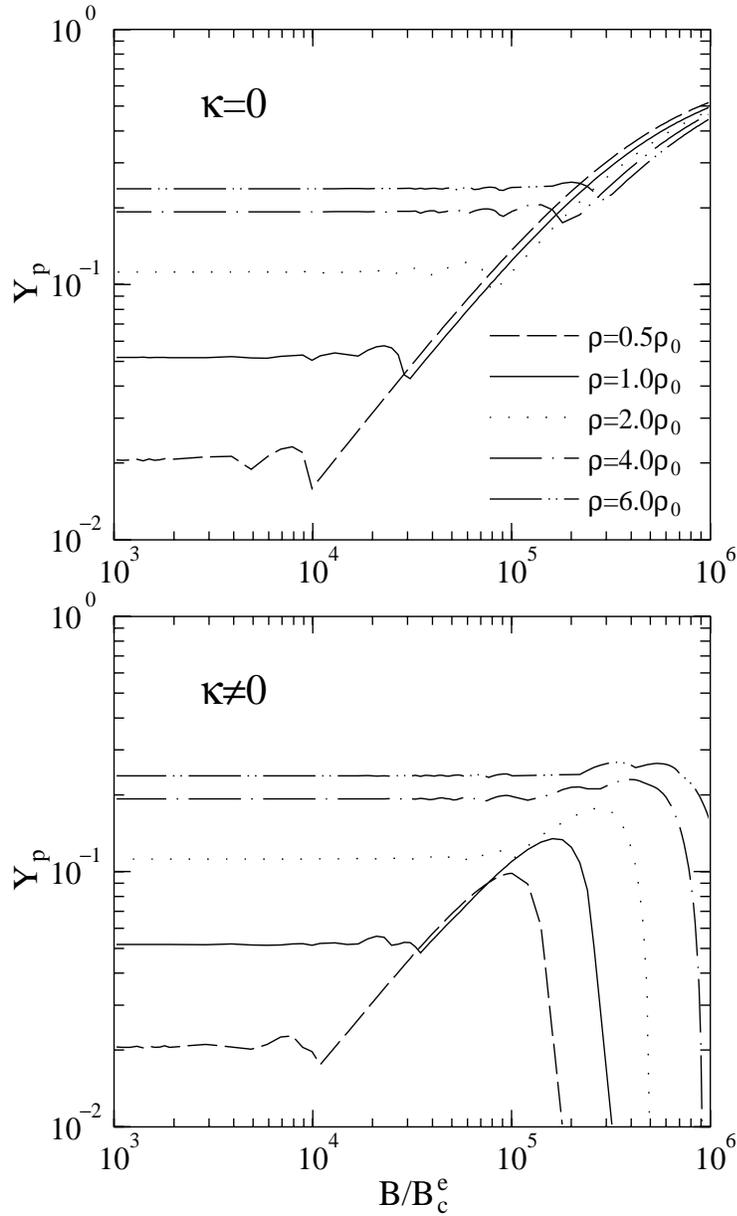,width=14.cm,height=18cm,angle=0}
 \vspace{0.5cm}
 \caption{The proton fraction as a function of the magnetic field strength.
 The upper panel shows the results without the AMM while in the lower
 panel the AMM is taken into account.}
 \label{polar}
\end{figure}

\begin{figure}[htbp]
  \vspace{4cm}
 \hskip  -2.0cm \psfig{file=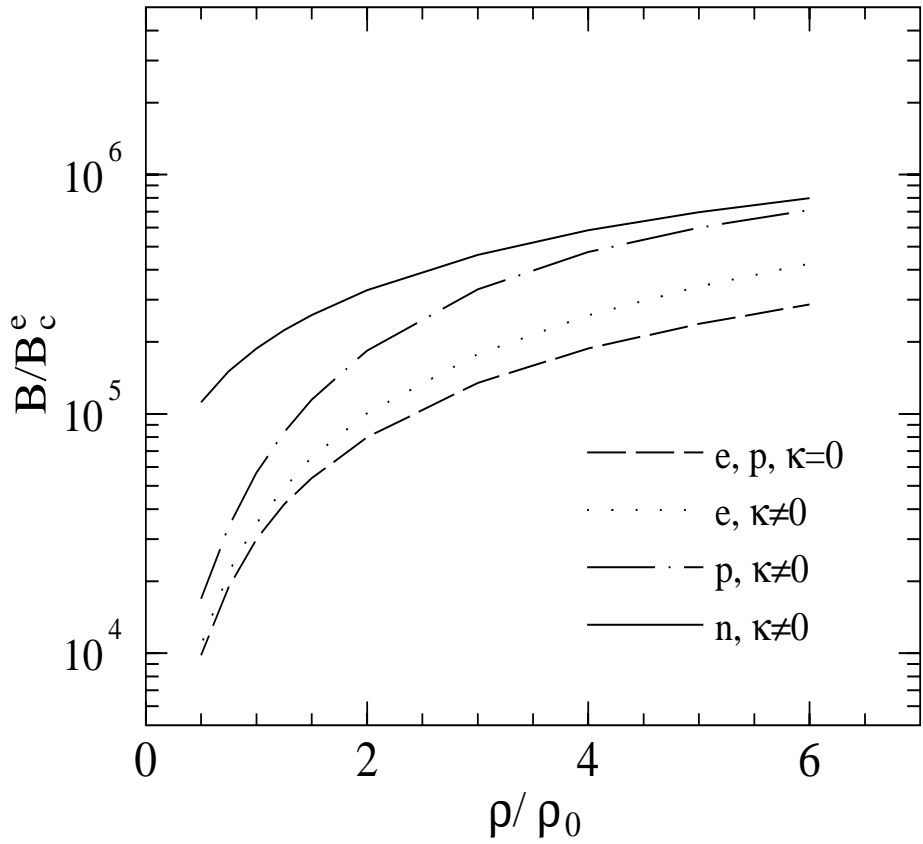,width=16.cm,height=15cm,angle=0}
 \vspace{-0.5cm}
 \caption{The critical field of particle spin polarization as a function of the
 baryon density. Different curves correspond to the different cases as
 indicated in the figure and discussed in the text.}
 \label{ratio}
\end{figure}

 In order to understand the above results, in the upper panel of Fig.~\ref{polar}
 we depict the proton fraction as a function of the magnetic field strength, where
 the effects of the AMM are neglected. It can be found that for a fixed
 density the proton fraction begins to increase rapidly at a certain value
 of the magnetic field. This is the critical point where both protons and electrons
 are completely spin polarized (i.e., only one
 Landau level is occupied.) due to Landau quantization
 while no direct effects from the magnetic field act on neutrons.
 The Fermi energies of electrons and protons fall drastically. As a consequence,
 a large amount of neutrons converts to protons. The system approaches to a
 well proton-rich matter. The variation of the critical field with the baryon
 density is displayed in Fig.~\ref{ratio} as the dashed line.

 In the lower panel of Fig.~\ref{polar} we show the results including the
 effects of the AMM. Different curves are related to the different baryon
 densities. For each curve there exist two turning points. The first one
 corresponds to the critical field of electron polarization, its variation
 with density is depicted as the dotted line in Fig.~\ref{ratio}. As soon as
 the electron is fully polarized, its Fermi energy will be considerably reduced.
 This causes the conversion of neutrons to protons in a beta-equilibrium
 system. The proton fraction is therefore enhanced substantially. Because
 of the different coefficients for the AMM of electrons and protons,
 the complete spin polarization of protons is reached later than
 electrons as shown in Fig.~\ref{ratio} with the dash-dotted line.
 There are no evident signals for the proton spin polarization in the lower
 panel of Fig.~\ref{polar} since the rapid increase of the proton fraction
 has been induced by the electron polarization. When the field is further
 increased, the neutron can be fully spin polarized because of the coupling of the
 neutron spin to the magnetic field, an effect totally caused by the
 anomalous magnetic moment. The inverse process, $p \rightarrow n$, starts
 at the second turning points appeared on the curves plotted in the lower
 panel of Fig.~\ref{polar}. This causes the Y$_{p}$ to drop rapidly with the
 increase of $B$. A pure neutron matter becomes possible at very large
 field strength. The critical field for neutron polarization as a function
 of the baryon density is given in Fig.~\ref{ratio} as the solid line.
 One can see that the neutron spin polarization happens at $B >
 10^{5}B_{c}^{e}$. We have also considered the situation where the
 electron AMM is switched off but the nucleon AMM is taken into account.
 No enhancement of the proton fraction is observed compared to the
 case that both the electron and nucleon AMM are neglected. While
 comparing with the condition that the AMM of both electrons and
 nucleons are considered, the proton fraction is less
 suppressed. For instance, protons start to appear at $\rho
 \approx 2 \rho_{0} $ rather than $\rho \approx 4\rho_{0}$ for
 $B = 10^{6}B_{c}^{e}$. In order to check our numerical process we
 have revealed the results of Ref. \cite{Bro00}, which is
 indicated in the lower panel of Fig.~1 as the long-dashed line.
 The above calculations have been repeated with other
 mean-field parameter sets used in Refs.~\cite{Cha97,Bro00}, the qualitative
 trend remains unchanged.

 In summary, we have studied the properties of neutron star matter under
 strong magnetic fields, with an emphasis on the neutron star composition.
 After taking into account the effects of the AMM of  nucleons and
 electrons, the proton fraction is found to never exceed the field free
 case. Our results demonstrate that extremely strong magnetic fields may lead to
 a pure neutron matter rather than a proton-rich matter observed in
 Refs. \cite{Cha97,Bro00}, mainly attributed to the complete
 neutron spin polarization induced by the AMM effects.

 \noindent {\bf Acknowledgments}:
 G.~Mao, V.N.~Kondratyev, Z.~Li, X.~Wu and I.N.~Mikhailov acknowledge the
 Japan Atomic Energy Research Institute for local hospitality
 during their visiting. G.~Mao thanks the financial support of the STA foundation.
 Z.~Li and X.~Wu acknowledge the National Natural Science
 Foundation of China for financial support.

 \end{sloppypar}

\newpage
 \newpage
 \newpage

\end{document}